\documentclass{INTERSPEECH2023}

\interspeechcameraready

\usepackage{multirow}
\usepackage{algorithm}  
\usepackage{algorithmic}
\usepackage{hyperref}
\usepackage{url}

\title{Complex Image Generation SwinTransformer Network for Audio Denoising}
\name{Youshan Zhang$^1$, Jialu Li$^2$}
\address{
  $^1$Yeshiva University, NYC, NY, USA\\
  $^2$Cornell University, Ithaca, NY, USA}
\email{youshan.zhang@yu.edu, jl4284@cornell.edu}

\begin{document}

\maketitle
 
\begin{abstract}
Achieving high-performance audio denoising is still a challenging task in real-world applications. Existing time-frequency methods often ignore the quality of generated frequency domain images.
This paper converts the audio denoising problem into an image generation task. We first develop a complex image generation SwinTransformer network to capture more information from the complex Fourier domain. We then impose structure similarity and detailed loss functions to generate high-quality images and develop an SDR loss to minimize the difference between denoised and clean audios. Extensive experiments on two benchmark datasets demonstrate that our proposed model is better than state-of-the-art methods. 
\end{abstract}

\section{Introduction}
Audio denoising aims to remove the background noise in the audio to generate better-quality information sources for real-life applications, such as speech enhancement~\cite{kong2021speech}, hearing aids~\cite{aubreville2018deep} and the lung~\cite{pouyani2022lung} and heart~\cite{kui2021heart} sounds for disease diagnosis. However, due to the degraded quality, unpleasant reverb, and loud background sound, pursuing high-quality denoised audio is still challenging. 


Deep learning methods have become prevalent in the audio denoising field, demonstrating a stronger ability to learn data features~\cite{wang2021denoising}.  In recent years, many time-frequency (T-F) domain deep-learning-based audio denoising approaches~\cite{kong2022speech} have been implemented using short-time Fourier transform (STFT) and applying inverse short-time Fourier transform (ISTFT) to denoise audios~\cite{wang2021tstnn}. Sonning et al.~\cite{sonning2020performance} investigated the performance of a time-domain network for speech denoising. The model was developed to deal with the original inability of STFT/ISTFT-based time-frequency approaches to capture short-time changes and was proved to be useful in a real-time setting. Wang et al.~\cite{wang2021tstnn} proposed a two-stage transformer neural network  for end-to-end audio denoising in the time domain. Their model included an encoder, a two-stage transformer module, a masking module, and a decoder, which outperformed many time- or frequency-domain models with less complex structures.

One problem in DNN-based audio denoising approaches is that they predict a label for each time frame from a small context window around the frame~\cite{tan2018convolutional}; therefore, it is difficult for models to track a target speaker among multiple interferences, which means that the DNNs are not easy to handle long-term contexts~\cite{li2020speech}. To cope with this problem, more deep learning approaches are proposed to better capture the audio features, e.g., recurrent neural networks (RNNs). Chen and Wang~\cite{chen2017long} proposed an RNN-based audio separation model with four hidden long short-term memory (LSTM) layers to deal with speaker generalization. Their model outperformed DNN-based models on unseen speakers and unseen noises regarding objective speech intelligibility. 
Maas et al.~\cite{maas2012recurrent} introduced a model using a deep recurrent auto-encoder neural network to denoise input features and capture the temporal nature of speech signals for robust automatic speech recognition~\cite{zhao2018convolutional}. 
 
To produce better noise audio processing results, Zhang et al.~\cite{zhang2016deep} built a novel deep recurrent convolutional network for acoustic modeling and then applied deep residual learning for audio recognition with faster convergence speed. Tan et al.~\cite{tan2018convolutional} proposed a recurrent convolutional network that incorporates a convolutional encoder-decoder and long short-term memory into the  convolutional recurrent neural network (CRN) architecture to address real-time audio enhancement. 
Li et al.~\cite{li2020speech}  combined the progressive learning framework with a causal CRN to further mitigate the trainable parameters and improve audio quality and intelligibility.  Zhang and Li~\cite{zhang2023birdsoundsdenoising} converted audio denoising into a visual image segmentation problem, and their results demonstrated that a better segmentation result leads to better audio denoising performance.

Audio denoising using waveform domain and transformer has also been explored. Kong et al.~\cite{kong2021speech} proposed an audio enhancement method with pre-trained audio neural networks using weakly labeled data and applied a convolutional U-Net to predict the waveform of individual anchor segments selected by PANNs. Kong et al.~\cite{kong2022speech} proposed CleanUNet, a causal speech denoising model on the raw waveform based on an encoder-decoder architecture combined with several self-attention blocks. Agarwal et al.~\cite{agarwal2022implementing} replaced the Bi-directional LSTM block with a transformer in the open-source Open-Unmix model for audio separation, and the new model trained faster than the unmodified model. However, these transformer methods only focused on denoising audios and usually did not check the quality of generated intermediate matrices.

\begin{figure*}[t]
  \centering
  \includegraphics[width=\linewidth]{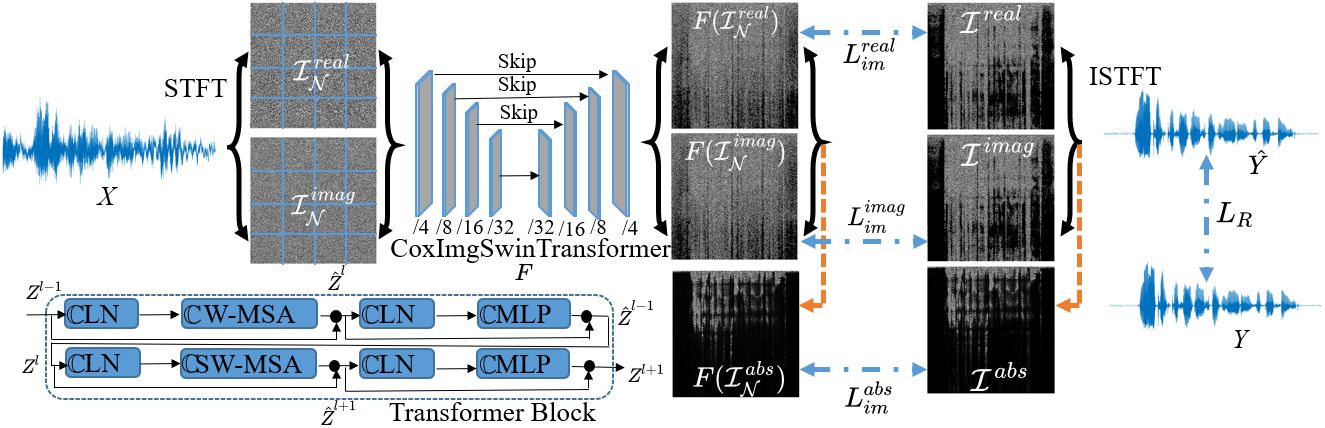}
  \vspace{-0.6cm}
  \caption{The schematic diagram of our CIGSN model. Each gray block is a transformer block in the CoxImgSwinTransformer module ($F$). We first apply STFT to convert audio signals $X$ into complex images (real image $\mathcal{I}_{\mathcal{N}}^{real}$ and imaginary image $\mathcal{I}_{\mathcal{N}}^{imag}$). Then, we feed them into the module ($F$), and get generated real $F(\mathcal{I}_{\mathcal{N}}^{real})$, imaginary $F(\mathcal{I}_{\mathcal{N}}^{imag})$, and absolute $F(\mathcal{I}_{\mathcal{N}}^{abs})$ images. Finally, we minimize the image quality check loss $L_{im}^{total} = L_{im}^{imag} + L_{im}^{real} + L_{im}^{abs}$, and audio reconstruction loss $L_R$.  }
  \label{fig:model}
  \vspace{-0.6cm}
\end{figure*}

To alleviate the aforementioned challenges, our contributions are three-fold:

\begin{itemize}
    \item We convert the audio denoising into an image generation problem. Our experiment demonstrates that a better-generated complex image will achieve better audio denoising performance.
    
    \item We develop a complex image generation SwinTransformer network (CIGSN) model, which is able to generate high-quality complex images in the Fourier domain.  

    \item We also propose image quality check and audio reconstruction modules. We enforce  image L1 loss, structural similarity loss, and detailed loss for the image quality check, and employee audio L1 loss and SDR loss to optimize the audio reconstruction module.

    
\end{itemize}

\section{Methodology}
In time domain audio denoising, a noisy audio signal $x$ can be typically expressed as:
\begin{equation}
  x = y  +  noise,  
\end{equation}
where $y$ and $noise$ denote clean audio and additive noise signal, respectively. Given noisy audio signals $X = \{x_i\}_{i=1}^{n}$, we aim to extract the clean audios $Y = \{y_i\}_{i=1}^{n}$ by learning a mapping $f$, and leverage $f(X) \approx Y$. In the Fourier frequency domain, we convert the audio denoising to an image generation task. Given the noisy audio complex images $\mathcal{I}_\mathcal{N} = \{{I_N}_i\}_{i=1}^{n}$ using $\text{STFT}(X)$ and clean audio complex images $\mathcal{I} = \{I_i\}_{i=1}^{n}$ using $\text{STFT}(Y)$,  we also aim to find a function $F$ such that $F(\mathcal{I}_\mathcal{N}) \approx \mathcal{I} $, where $F(\mathcal{I}_\mathcal{N})$ is the generated complex images.

\vspace{-0.2cm}
\subsection{Motivation}
The existing time-frequency audio denoising methods majorly convert the audio signal to the Fourier domain using STFT and get the reconstructed matrix, and then apply the ISTFT to get the denoised audio. However, the intermediate process of the reconstructed matrix is usually less explored. We aim to pursue a high-quality generated matrix (complex images) and convert it to an image generation problem in the Fourier domain. 

\vspace{-0.2cm}
\subsection{CoxImgSwinTransformer}
To generate high-quality real and imaginary images, we develop a CoxImgSwinTransformer model. The details of SwinTransformer can be found in~\cite{liu2021swin}. However, the original SwinTransformer model cannot handle complex image inputs. We hence develop a complex image inputs variant of the SwinTransformer model (CoxImgSwinTransformer).  Given an input batch of tensor $T = [N, C, H, W]$, where $N$ is the number of samples in the batch; $C$ is the channel size ( $C =1$ if the audio is a single track, and $C =2$ if the audio is dual tracks); $H$ and $W$ are the height and width of the image (note that the tensor $ T = A + jB \in \mathbb{C} ^{N \times C \times H \times W} $ is a complex number, $A$ is the real part, and $B$ is the imaginary part of the complex tensor),  we define the basic deep learning operations $O$ as: 
\begin{equation}\label{eq:o}  
    \mathbb{C}O (T) = O (A) + jO(B),
\end{equation}
where $j$ is the square root of $-1$ and $O$ can be common deep learning layers (Conv2d, MaxPool2d, BatchNorm2d, ReLU, GeLU, Dropout, Interpolate, Sigmoid, LayerNorm, Softmax, Linear, etc.). By applying Eq.~\eqref{eq:o}, we can get the complex version of these layers as ($\mathbb{C}$Conv2d, $\mathbb{C}$MaxPool2d, $\mathbb{C}$BatchNorm2d, $\mathbb{C}$ReLU, $\mathbb{C}$GeLU, $\mathbb{C}$Dropout, $\mathbb{C}$Interpolate, $\mathbb{C}$Sigmoid, $\mathbb{C}$LayerNorm, $\mathbb{C}$Softmax, $\mathbb{C}$Linear, etc.). With the basis of these layers, we can build the CoxImgSwinTransformer model. Fig.~\ref{fig:model} shows the overall architecture of our proposed CoxImgSwinTransformer module, which has three key parts: encoder, decoder, and skip connections. 

\subsubsection{Encoder}
The encoder consists of four Swin transformer blocks. Each Swin transformer block is composed of a complex attention layer and a complex feed-forward layer, including a complex LayerNorm ($\mathbb{C}$LN) layer, complex multi-head self-attention module, a two-fully connected layers complex MLP ($\mathbb{C}$MLP), and a $\mathbb{C}$GELU nonlinearity layer.  The $\mathbb{C}$LN and $\mathbb{C}$GELU are computed based on Eq.~\eqref{eq:o}. 

The $\mathbb{C}$MLP module has five complex layers: $\mathbb{C}$Linear, $\mathbb{C}$GELU, $\mathbb{C}$Dropout, $\mathbb{C}$Linear and $\mathbb{C}$Dropout. Between two successive transformer blocks, there is a complex window-based multi-head self-attention ($\mathbb{C}$W-MSA) module, and a complex shifted window-based multi-head self-attention ($\mathbb{C}$SW-MSA) module. The continuous swin transformer process is represented as: 
\begin{equation}
\begin{aligned}
& \hat{Z}^{l} = \mathbb{C}W\text{-}MSA(\mathbb{C}LN(Z^{l-1})) + Z^{l-1} \\
& Z^{l} =  \mathbb{C}MLP ( \mathbb{C}LN (\hat{Z}^{l})) + \hat{Z}^{l} \\
& \hat{Z}^{l+1} = \mathbb{C}SW\text{-}MSA(\mathbb{C}LN(Z^{l})) + Z^{l} \\
& Z^{l+1} =  \mathbb{C}MLP ( \mathbb{C}LN (\hat{Z}^{l+1})) + \hat{Z}^{l+1}, \\
\end{aligned}
\end{equation}
where $\hat{Z}^{l}$ and $Z^l$ represent the outputs of the $\mathbb{C}(S)$W-MSA module and the $\mathbb{C}$MLP module of the $l^{th}$ block, respectively. The complex  self-attention is computed according to:
\begin{equation}
\mathbb{C}Attention(\mathbb{C}Q,\mathbb{C}K,\mathbb{C}V)=\mathbb{C}SoftMax(\frac{\mathbb{C}Q\mathbb{C}K^{\mathbb{C}T}}{\sqrt{d}} +B)\mathbb{C}V,
\end{equation}
where $\mathbb{C}Q, \mathbb{C}K, \mathbb{C}V\in \mathbb{C}^{M^2\times d}$ are the query, key and value matrices; d is the query/key dimension, $M^2$ is the number of patches in a window and $B$ is taken from bias matrix $\hat{B} \in \mathbb{C}^{{(2M-1)\times (2M+1)}}$.


In the encoder, the dimensions of features in four transformer blocks are $\frac{H}{4} \times \frac{W}{4} \times C$, $\frac{H}{8} \times \frac{W}{8} \times 2C$, $\frac{H}{16} \times \frac{W}{16} \times 4C$, $\frac{H}{32} \times \frac{W}{32} \times 8C$, which corresponds to $/4, /8, /16,$ and $/32$ of Fig.~\ref{fig:model}, respectively. For the patch merging layer, we concatenate the input features of each group of 2 $\times$ 2 neighboring patches and use the $\mathbb{C}$linear layers to obtain the specified channel number of output features.  

\vspace{-0.2cm}
\subsubsection{Decoder}
In the  decoder, we also have four symmetric transformer blocks. However, we use the patch expanding layer in the decoder to upsample the extracted deep features. The output dimension of the four blocks are: $\frac{H}{32} \times \frac{W}{32} \times 8C$, $\frac{H}{16} \times \frac{W}{16} \times 4C$, $\frac{H}{8} \times \frac{W}{8} \times 2C$, and $\frac{H}{4} \times \frac{W}{4} \times C$, which corresponds to $/32, /16, /8,$ and $/4$ of Fig.~\ref{fig:model}, respectively. 

\vspace{-0.2cm}
\subsubsection{Skip connection}
We applied three skip connections to fuse the multi-scale features from the encoder with the decoder. We concatenate the shallow features from the encoder and deep features from the decoder to reduce spatial information loss and form robust features. The final output dimensions of height and width are the same as the input images. 

\vspace{-0.2cm}
\subsection{Image quality check}\label{sec:img}
Given complex input images $\mathcal{I}_\mathcal{N}$, our CoxImgSwinTransformer model can output the generated complex images $F(\mathcal{I}_\mathcal{N})$. 
To improve generated image quality, we developed an image quality check module. We first apply L1 loss to minimize the difference between the generated images $F(\mathcal{I}_\mathcal{N})$ and the ground truth images $\mathcal{I}$, $L_{F} = |F(\mathcal{I}_\mathcal{N}) - \mathcal{I}|_{1}$, where $F$ is our proposed CoxImgSwinTransformer, $\mathcal{I}_\mathcal{N}$ is noise complex images. Secondly, to further improve the generated images, we impose a structural similarity loss $L_{S}$ to examine the generated image quality. The $L_{S}$ is defined as: $ 1 - SSIM(F(\mathcal{I}_\mathcal{N}), \mathcal{I})$, where $SSIM$ is the structural similarity~\cite{wang2004image}. The range of the $L_S$ is from 0 to 1, where 0 indicates high similarity between images and 1 means they are not similar. Finally, to enhance image details, we add a detailed loss, called $L_{D}$ and it is given by:
\begin{equation}
    L_{D} = |VGG(F(\mathcal{I}_\mathcal{N}))- VGG(\mathcal{I}) |_1,
\end{equation}
where VGG is the activation of the last fully connected layer from the pre-trained VGG19 network. 
Therefore, we improve the quality of the generated images by minimizing Eq.~\eqref{eq:im}.
\begin{equation}\label{eq:im}
    L_{im} = L_{F} + L_{S} + L_{D} 
\end{equation}
As shown in Fig.~\ref{fig:model}, we could get three different images given one audio: the real image, the imaginary image, and the absolute image ($abs$ takes the absolute value of a complex tensor from the output of CoxImgSwinTransformer). Eventually, the image quality check module loss consists of these three image minimizations and is defined as:
\begin{equation}\label{eq:loss_image}
    L_{im}^{total} = L_{im}^{imag} + L_{im}^{real} + L_{im}^{abs},
\end{equation}
where $L_{im}^{imag} = L_{F}^{imag} + L_{S}^{imag} + L_{D}^{imag}$, and similarly, we could change $imag$ of $L_{im}^{imag}$ into $real$, and $abs$ to get $L_{im}^{real}$ and $L_{im}^{abs}$, respectively. 

\vspace{-0.2cm}
\subsection{Audio reconstruction}\label{sec:audio}
After getting the output from the decoder layers from the CoxImgSwinTransformer model, we could apply ISTFT to get the reconstructed audio as $\hat{Y}$. We first apply L1 loss to minimize the difference between reconstructed audio and the ground truth as $L_{A} = |\hat{Y}-Y|_1$. We also propose an SDR loss to evaluate the quality of $\hat{Y}$. The SDR is defined as: $SDR(\hat{Y}, Y) = 10\  \text{log}_{10}\frac{||Y||^2}{||\hat{Y} - Y||^2}$. We defined the SDR loss as:
\begin{equation}
    L_{SDR} = const_{upper}- SDR(\hat{Y}, Y),
\end{equation}
where $const_{upper}$ is the upper bound constant value, we set it as 20. Therefore, we could ensure that the SDR loss keeps decreasing during the training. The audio reconstruction loss is defined as:
\begin{equation}\label{eq:loss_audio}
    L_{R} = L_{A} + L_{SDR}.
\end{equation}

\vspace{-0.2cm}
\subsection{Objective Function}
The architecture of our proposed CoxImgSwinTransformer model is shown in Fig.~\ref{fig:model}. Considering all loss functions in Sec.~\ref{sec:img} and Sec.~\ref{sec:audio}, our model minimizes the following objective function
\begin{equation}\label{eq:overall_loss}
    L = \alpha L_{im}^{total} + (1-\alpha) L_{R},
\end{equation}
where $\alpha$ is the balance factor between all image loss and audio reconstruction loss. This objective function enables us first to get high-quality generated images and then acquire a better reconstructed denoised audio. 
Our training procedures are described in Alg.\ref{alg:vision}. 

\vspace{-0.2cm}
\begin{algorithm}[h]
   \caption{Complex Image Generation SwinTransformer Network (CIGSN). $B(\cdot)$ denotes the mini-batch training sets, $L$ is the number of iterations.}
   \label{alg:vision}
\begin{algorithmic}[1]
   \STATE {\bfseries Input:} Noisy audio signals  $X = \{x_i\}_{i=1}^{n}$ and clean audio signals  $Y = \{y_i\}_{i=1}^{n}$, where $n$ is the number of audios.
   \STATE {\bfseries Output:} Denoised audio signals
   \STATE Generate noise audio images $\mathcal{I}_\mathcal{N} = \{{I_N}_i\}_{i=1}^{n}$ and clean audio images $\mathcal{I} = \{I_i\}_{i=1}^{n}$ using STFT
   \FOR{$iter =1$ {\bfseries to} $L$}
   \STATE Derive $B(\mathcal{I}_\mathcal{N})$ and $ B(\mathcal{I})$ sampled from $\mathcal{I}_\mathcal{N}$ and $\mathcal{I}$
   \STATE Calculate image quality check loss using Eq.~\eqref{eq:loss_image}
   \STATE Convert complex images to audio signals using ISTFT and calculate audio reconstruction loss using Eq.~\eqref{eq:loss_audio}
   \STATE Optimize CIGSN model $F$ using Eq.~\eqref{eq:overall_loss}
   \ENDFOR
   \STATE Output $F(\mathcal{I}_\mathcal{N})$, and get denoised audios using ISTFT
\end{algorithmic}
\end{algorithm}
\vspace{-0.6cm}

\section{Experiments}
\vspace{-0.1cm}
\subsection{Datasets}
We evaluate our model using two benchmark datasets.

\noindent \textbf{VoiceBank-DEMAND}~\cite{valentini2017noisy} is a synthetic dataset created by mixing up clean speech and noise.
The training set contains 11,572 utterances (9.4h), and
the test set contains 824 utterances (0.6h). The lengths of utterances range from 1.1s to 15.1s, with an average of 2.9s.

\noindent \textbf{BirdSoundsDenoising}~\cite{zhang2023birdsoundsdenoising} contains 14,120 audios and is a large-scale dataset of bird sounds collected containing 10000/1400/2720 in training, validation, and testing, respectively. Unlike many audio-denoising datasets, which have manually added artificial noise, these datasets contain many natural noises, including wind, waterfall, rain, etc.

\begin{table}[h!]
  \caption{Comparison results on the VoiceBank-DEMAND dataset. ``$-$" means not applicable.}
  \label{tab:voice}
  \centering
  \vspace{-0.3cm}
\setlength{\tabcolsep}{+0.3mm}{
  \begin{tabular}{lcllllll}
    \toprule
    Methods     & Domain & PESQ & STOI & CSIG & CBAK & COVL &SSIM \\
    \hline
CP-GAN~\cite{liu2020cp} &T &2.64 &0.942 &3.93& 3.33 &3.28& 0.58 \\
PGGAN~\cite{li2022perception} &T &2.81 &0.944& 3.99& 3.59 &3.36& 0.56 \\
DCCRGAN~\cite{huang2022dccrgan} &TF &2.82 &0.949 &4.01& 3.48 &3.40 & 0.65\\
S-DCCRN~\cite{lv2022s} &TF &2.84 &0.940 &4.03 &2.97 &3.43 & 0.62\\
DCU-Net~\cite{choi2019phase} &TF &2.93 &0.930 &4.10 &3.77& 3.52 & 0.67\\
PHASEN~\cite{yin2020phasen} &TF &2.99 &$-$ &4.18& 3.45& 3.50 & 0.72\\
MetricGAN+~\cite{fu2021metricgan+} &TF &3.15 &0.927& 4.14& 3.12 &3.52 & 0.78 \\
TSTNN~\cite{wang2021tstnn} & T & 2.96 & 0.950 & 4.33 & 3.53 & 3.67 & 0.78\\
MANNER~\cite{park2022manner} & T & 3.21 & 0.950 & 4.53 & 3.65 & 3.91 & 0.81\\
\hline
CIGSN & TF & \textbf{3.41} & \textbf{0.954} & \textbf{4.78} & \textbf{3.82} & \textbf{4.22} & \textbf{0.88}\\
\hline
  \end{tabular}}
  \vspace{-.7cm}
\end{table}

\begin{figure*}[t]
  \centering
  \includegraphics[width=0.8  \linewidth]{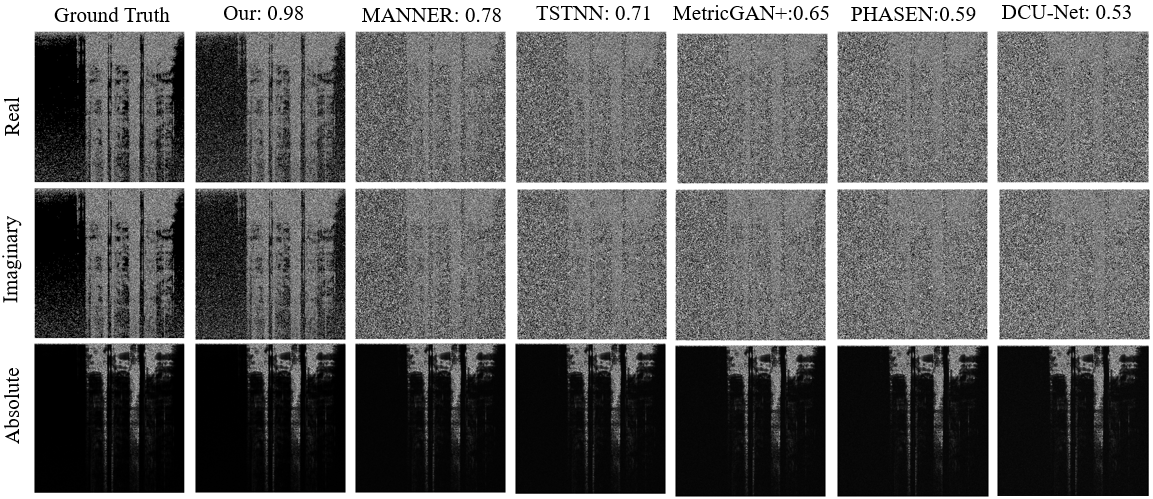}
  \vspace{-0.3cm}
  \caption{Comparison of generated images of our CIGSN and the best five baseline methods of real, imaginary, and absolute images.   }
  \label{fig:re}
    \vspace{-0.6cm}
\end{figure*}

\vspace{-0.2cm}
\subsection{Implementation details}\label{sec:implement}
During the training, we set batch size = 16, training iteration $L = 100$, and learning rate = 0.001, $\alpha = 0.5$ with an Adam optimizer on a 48G RTX A6000 GPU using PyTorch. We applied the STFT to convert audio signals to audio images and utilized 1000-point Hamming as the window function, the size of Fourier transform $n\_fft = 1023$. The length of each audio can be different, and we set the distance between neighboring sliding window frames as $hop\_length=int(length(y_t)/512)$, where $length(y_t)$ is the length of each audio. We then resize the input image dimensions as $[512 \times 512 \times 1]$\footnote{Source code is available at \url{https://github.com/YoushanZhang/CoxImgSwinTransformer.}}.

\vspace{-0.2cm}
\subsection{Results}
Tab.~\ref{tab:voice} shows the comparison results of the VoiceBank-DEMAND dataset. For nine baseline models, we also get the generated images following the same STFT parameters in Sec.~\ref{sec:implement}. We reported the extra structure similarity (SSIM) between generated image and the ground truth (mean SSIM of real, imaginary, and absolute images). Our proposed CIGSN model achieves the highest performance in all six metrics. Particularly, the SSIM metric is much higher than all other methods. To explore the underlying reason, we compare generated complex images with the five best baselines as shown in Fig.~\ref{fig:re}. We also list their mean SSIM score. The generated real and imaginary images of the CIGSN model are close to ground truth, while the other five models contain many noise areas. Surprisingly, the absolute images of the other five methods are similar to ground truth, which is caused by absolute images taking the absolute values of real and imaginary images so that there are no visible negative values. We find that a higher SSIM score (better-generated images) achieves better audio denoising performance. Tab.~\ref{tab:md} presents the results of the BirdSoundsDenoising dataset. Results of F1, IoU, and Dice are omitted since these metrics are used for the audio image segmentation task~\cite{zhang2023birdsoundsdenoising}. Our CIGSN still outperforms all other models in terms of SDR.

\begin{table}[t]
\small
\begin{center}
\captionsetup{font=small}
\caption{Comparison results of the BirdSoundsDenoising dataset ($F1, IoU$, and $Dice$ scores are multiplied by 100.  }
\vspace{-0.3cm}
\setlength{\tabcolsep}{+0.3mm}{
\begin{tabular}{lllll|lllllllll}
\hline \label{tab:md}
 \multirow{2}{*}{Networks}
 &  \multicolumn{4}{c}{Validation} & \multicolumn{4}{c}{Test} \\
 \cmidrule{2-9}
& $F1$ & $IoU$  & $Dice$ & $SDR$ & $F1$ & $IoU$  & $Dice$ & $SDR$ \\
\hline
MTU-NeT~\cite{wang2022mixed}  &69.1 &56.5 &69.0  &8.17 & 68.3  &55.7 & 68.3 &7.96  \\
Segmenter~\cite{strudel2021segmenter} & 72.6  & 59.6 & 72.5 & 9.24 & 70.8 & 57.7 & 70.7 & 8.52   \\
U-Net~\cite{ronneberger2015u}  &75.7 &64.3 &75.7 & 9.44 &74.4 &62.9 &74.4 & 8.92    \\
SegNet~\cite{badrinarayanan2017segnet}  &77.5 &66.9 &77.5 & 9.55&76.1 &65.3 &76.2 & 9.43 \\
DVAD~\cite{chen2018encoder}  & 82.6  & 73.5 & 82.6 &  10.33  & 81.6 & 72.3 & 81.6 & 9.96 \\
R-CED~\cite{park2017fully} & $-$ & $-$ & $-$ &2.38     &$-$ &$-$&$-$ & 1.93  \\
Noise2Noise~\cite{kashyap2021speech}  & $-$ & $-$ & $-$ & 2.40&$-$ &$-$&$-$ &1.96\\
TS-U-Net~\cite{moliner2022two}  & $-$ & $-$ & $-$ & 2.48&$-$ &$-$&$-$ &1.98\\
\hline
CIGSN  & $-$ & $-$ & $-$ & \textbf{10.69}&$-$ &$-$&$-$ &\textbf{10.15}\\
\hline
\end{tabular}}
\end{center}
\vspace{-.8cm}
\end{table}


\begin{table}[h!]
  \caption{Ablation study of different modules}
  \label{tab:ab}
  \centering
  \vspace{-0.3cm}
\setlength{\tabcolsep}{+1.5mm}{
  \begin{tabular}{lcllllll}
    \toprule
    Methods     & U+I & U+A & U+I+A & C+I & C+A & C+I+A\\
    \midrule
SDR&8.54 &7.97 & 8.98 & 10.0 & 9.84 &10.69 \\
\bottomrule
  \end{tabular}}
  \vspace{-0.6cm}
\end{table}

\noindent \textbf{Ablation study} To demonstrate the effectiveness of the proposed three modules: CoxImgSwinTransformer, image quality check, and audio reconstruction, we conduct an ablation study using the BirdSoundsDenoising validation dataset in Tab.~\ref{tab:ab}. ``U" means the complex U-Net model~\cite{choi2019phase}, ``I" means image quality check module, ``A" means audio reconstruction module and ``C" means CoxImgSwinTransformer model. Our CoxImgSwinTransformer model is better than the complex U-Net model. In addition, the image quality check module is more important than the audio reconstruction module. 

From the above experiments,  we can conclude that our proposed CIGSN model is effective in audio-denoising tasks. There are two compelling reasons. Firstly,  our CoxImgSwinTransformer module can distill real and imaginary images, and we can directly visualize the generated complex images. Secondly, the proposed image check and audio reconstruction module is able to minimize the models' prediction and ground truth. One weakness of the model is that it requires high GPU memory to train the CoxImgSwinTransformer module.

\vspace{-0.3cm}
\section{Conclusions}
In this paper, we convert the audio denoising into an image generation problem. We first develop a complex image generation SwinTransformer network to capture more information from the complex images. We then impose structure similarity loss to generate high-quality images and develop an SDR loss to minimize the difference between denoised audio and clean audio. Extensive experiments demonstrate our proposed model outperforms state-of-the-art models.

\bibliographystyle{IEEEtran}
\bibliography{mybib}

\end{document}